\documentclass[preprint2]{aastex62}
\usepackage[utf8]{inputenc}
\usepackage{graphicx}
\usepackage{wrapfig}
\usepackage{array}
\usepackage{tabularx}
\usepackage{lineno}
\usepackage{amsmath}
\usepackage{xcolor}

\begin{document}

\title{Changes in a Dusty Ringlet in the Cassini Division after 2010}
\author{Mathew Hedman}
\affil{Department of Physics, University of Idaho, Moscow Idaho 83843}
\author{Bill Bridges}
\affil{Department of Physics, University of Idaho, Moscow Idaho 83843}

\begin{abstract}
A dusty ringlet designated R/2006 S3, also known as the ``Charming Ringlet”, is located around 119,940 km from the center of Saturn within the Laplace Gap in the Cassini Division. Prior to 2010, the ringlet had a simple radial profile and a predictable eccentric shape with two components, one forced by solar radiation pressure and the other freely precessing around the planet. However, observations made by the Cassini spacecraft since late 2012 revealed a shelf of material extending inwards from the ringlet that was not present in the earlier observations. Closer inspection of images obtained after 2012 shows that sometime between 2010 and 2012 the freely-precessing component of the ringlet's eccentricity increased by over 50\%, and that for at least 3 years after 2012 the ringlet had longitudinal brightness variations that rotated around the planet at a range of rates corresponding to $\sim$60 km in orbital semi-major axis. Some event therefore disturbed this ringlet between 2010 and late 2012.

\bigskip

\bigskip
\end{abstract}

\vspace{2cm}

\section{Introduction}

Saturn's complex ring system includes multiple broad rings and narrow ringlets that are primarily composed of particles less than 100 microns across. Unlike the millimeter-to-meter sized particles that dominate the main ring system, these dusty rings are sensitive to a variety of non-gravitational forces that can influence their structure and dynamics. Furthermore, several of these dusty rings have been observed to change significantly over timescales of years to decades.  Most often these changes  include the formation of bright clumps of material  \citep{French12, Hedman13, Hedman19}, but in other cases they involve larger scale structural changes that can be attributed to variations  in periodic perturbing forces or discrete disturbances spanning broad ring regions \citep{Chancia19, Hedman16}. This paper describes a new type of time-variable phenomena in dusty rings, where the radial profile and orbit shape of a narrow ringlet appears to have suddenly changed.

The particular ringlet we will focus on here is located within the Laplace Gap of the Cassini Division in Saturn's rings, and is officially designated R/2006 S3 \citep{Porco06}, but informally referred to as the ``Charming Ringlet". The ringlet has a peak optical depth of $10^{-3}$ and is strongly forward-scattering, indicating that it is composed primarily of dust-sized particles \citep{Horanyi09, Hedman11f}. Observations from early in the Cassini mission showed that this ringlet exhibited heliotropic behavior (i.e. the ringlet's peak brightness is found further from the planet at longitudes more closely aligned to the Sun), which is most likely due to solar radiation pressure perturbing the orbits of the small particles that form this ringlet \citep{Hedman10}. 

\begin{figure}
    \centering
    \includegraphics[width=3in]{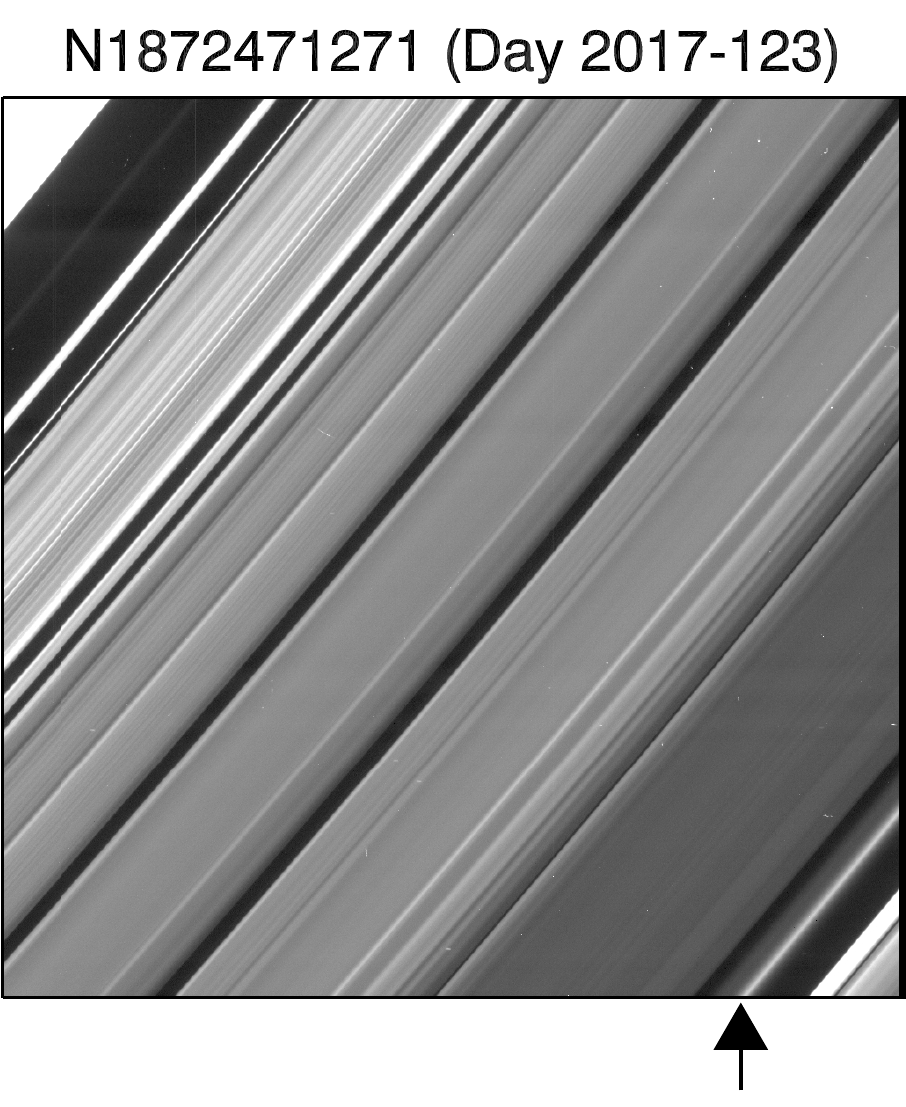}
    \caption{{The highest resolution and signal-to-noise image of the ringlet obtained in 2017 at a phase angle of 152$^\circ$.  Radius increases from upper left to lower right, and the ringlet's location is indicated by the arrow. This image shows a clear ``shelf" of material extending from the ringlet's inner flank that was not present in earlier images (see Figure~\ref{shelfim2}).}} 
    \label{shelfim}
\end{figure}

\begin{figure*}
    \rotatebox{90}{\includegraphics[width=3.7in]{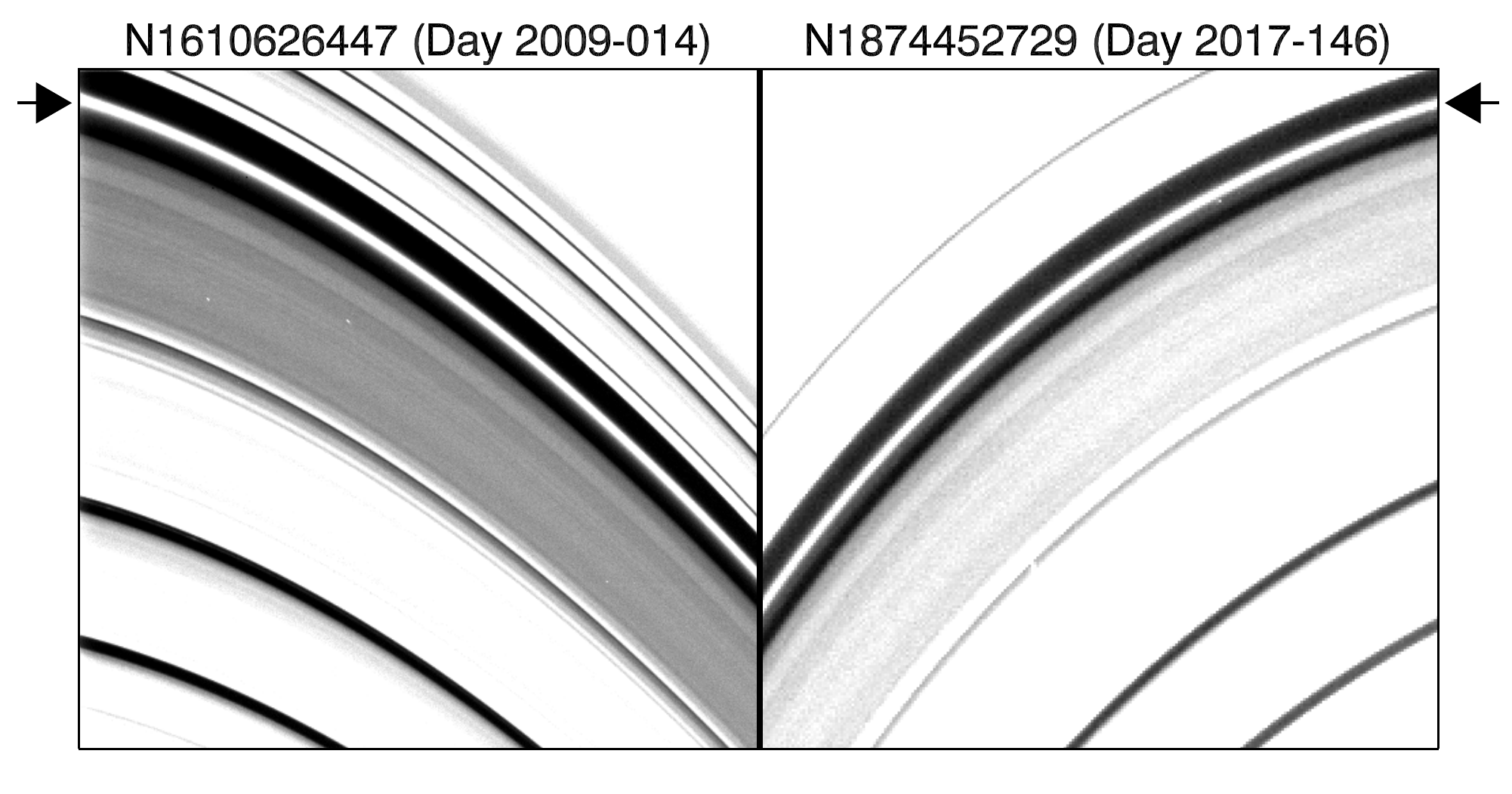}}
    \caption{{Changes in the morphology of the dusty ringlet within the Laplace gap. The two images shown here were both obtained at phase angles of 143$^\circ$ and have been independently rotated, cropped to facilitate comparisons. The ringlet is marked by the arrows. The left image was obtained in 2009 while the right image was obtained in 2017 (see Figure~\ref{profsamp} for radial brightness profiles derived from these images). Note that in the earlier image the ringlet appears as a relatively symmetric bright band, while in the later image there is again a faint  ``shelf" of material extending from the ringlet's inner flank.} }
    \label{shelfim2}
\end{figure*}

Images obtained later in the Cassini mission reveal a noticeable change in this ringlet's radial profile. Prior to 2010, the ringlet's brightness declined roughy symmetrically around its peak, but after 2012  a ``shelf'' of material can be seen on its inner flank. {Figure~\ref{shelfim} shows the best image of this feature, while Figure~\ref{shelfim2} shows two images of the Charming Ringlet taken at similar phase angles eight years apart, where the shelf is  visible in the later image and is absent from the earlier image.} Similarly, Figure~\ref{profsamp} shows radial brightness profiles of the ringlet obtained before and after 2010 at nearly identical phase angles, demonstrating that the shelf is seen repeatedly in images taken after 2012, and is always absent from earlier observations. These data therefore indicate that something happened to this ringlet over the course of the Cassini mission that changed its overall structure. 
 
In this paper we use the full span of the Cassini data to quantify and characterize these changes to the ringlet, and to explore what processes might be responsible for disturbing this ringlet.  Section~\ref{obs} describes the Cassini imaging data used in this study and how these data were processed to generate radial brightness profiles of this ringlet. Section~\ref{history} then examines both the total brightness of the ringlet and its radial position to show that while the particle content of the ringlet did not change dramatically over the course of the Cassini mission, its shape and orbital properties underwent a clear shift between 2010 and 2012. Section~\ref{long} takes a closer look at the brightness profiles obtained after 2010, which reveal that the formation of the shelf was associated with significant longitudinal variations in the ringlet's brightness that rotated around the planet at speeds consistent with the expected mean motion of orbiting material. Finally, Section~\ref{discussion} discusses the processes that could have been responsible for inducing this change, which include collisions with interplanetary debris and sudden changes in the ring's electromagnetic environment. 

 
\section{Observations and preliminary data reduction}
\label{obs}

This study uses images obtained by the  Narrow Angle Camera of the Image Science Subsystem (ISS) onboard the Cassini Spacecraft \citep{Porco04}. We conducted a comprehensive search of images containing the Laplace Gap with resolutions better than 10 km and ring opening angles above 5$^\circ$ using the OPUS search tool available on the Ring-Moon Systems Node of the Planetary Data System ({\tt  https://pds-rings.seti.org/search}).\footnote{Later searches found a few additional image sequences that contained the ringlet, but these were judged to have too low signal-to-noise and/or resolution for this particular study.} This search yielded 1228 images. A relatively small number of the above images were found not to be suitable for this analysis. First of all, we excluded 93 images that were made when the Sun was within 2$^\circ$ of the ringplane because during this timeframe shadows from nearby ring material could fall across the ringlet, complicating the analysis. A further 17 images were removed from consideration because the ringlet was either not completely captured in the image or the image could not be properly navigated. These issues caused the total radially-integrated brightness of the ringlet to be extremely high or low (above  100 m or below 0.1 m) or the apparent ringlet position to be over 80 km from its expected location. This left a total of 1118 images deemed suitable for this particular study.

\begin{figure}
\resizebox{3.25in}{!}{\includegraphics{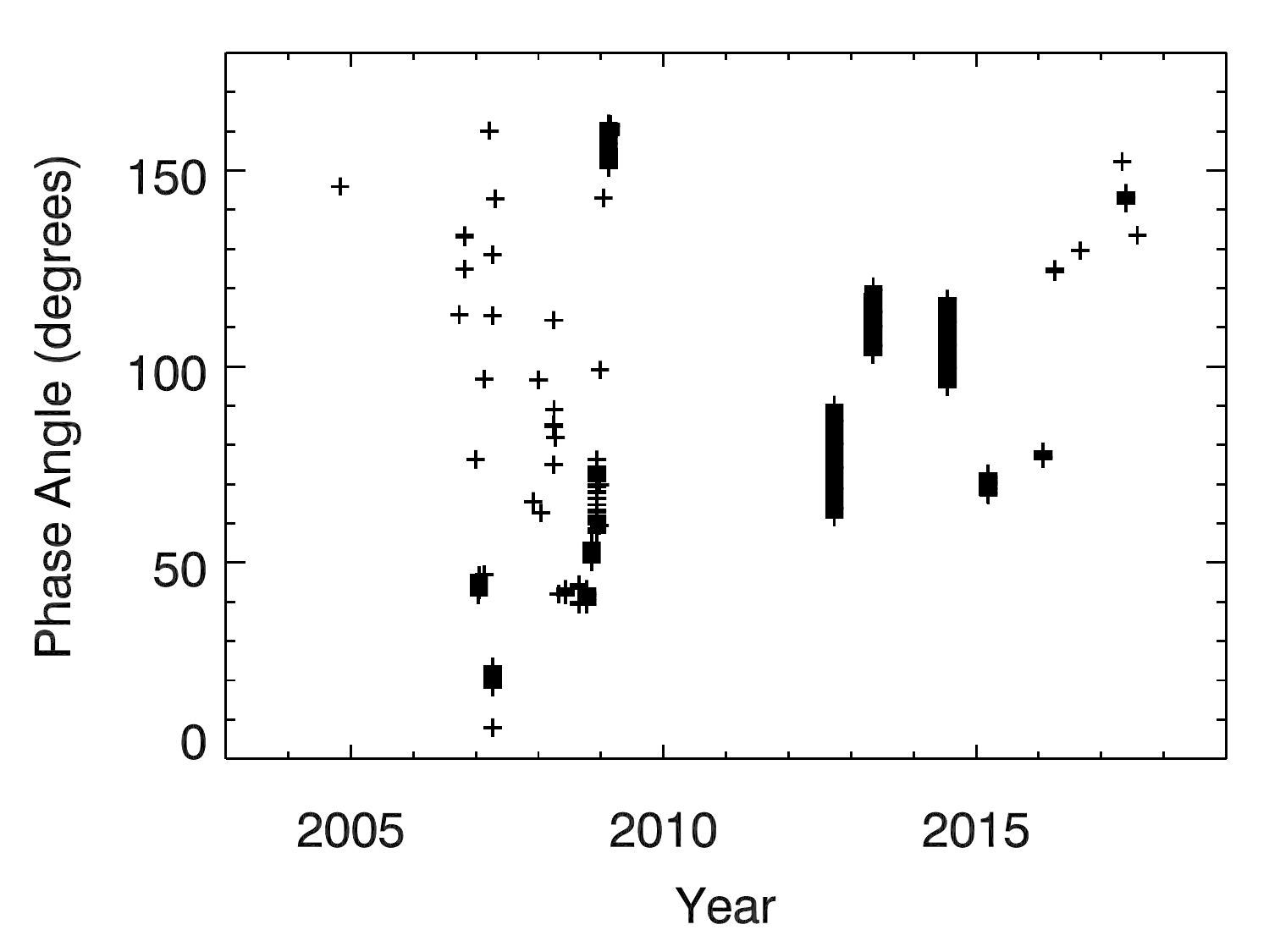}}
\caption{Distribution of the ringlet observations used in this analysis. Each point corresponds to  the time and phase angle of an individual image. Note the large data gap between late 2009 and late 2012. Also note that the data prior to 2010 contain a relatively large fraction of isolated observations, while the data obtained after 2012 is mostly in the form of a few sequences of many images.}
\label{datpars}
\end{figure}

\begin{table*}
\begin{center}
    \caption{Notable movie sequences of the Charming Ringlet}
    \label{obstab}
    \resizebox{6.5in}{!}{\hspace{-1in}%
    \begin{tabular}{|c|c|c|c|c|c|}
    \hline
    Observation ID & Files & Date  & Duration (hours) & Phase Angle (deg) & Emission Angle (deg) \\
    \hline
    ISS\textunderscore103RI\textunderscore SHRTMOV001\textunderscore PRIME & N1613254649-N1613295800 & 2009-044 & 11 & 161.897-150.715 & 96.8-116.18 \\
    \hline
    ISS\textunderscore104RI\textunderscore SHRTMOV002\textunderscore PRIME & N1614278585-N1614302748 &  2009-056 & 6 & 161.54-159.198 & 93.13-104.57 \\
    \hline
    ISS\textunderscore172RI\textunderscore MOONLETCD001\textunderscore PIE & N1727207218-N1727222674 &  2012-268 & 4 & 90.207-61.55 & 56.36-58.03 \\
    \hline
    ISS\textunderscore189RI\textunderscore BMOVIE001\textunderscore PRIME & N1746791718-N1746835578 &  2013-129 & 12 &  103.14-120.27 & 148.22-152.91 \\
    \hline
    ISS\textunderscore206RI\textunderscore BMOVIE001\textunderscore PRIME & N1784298322-N1784343322 &  2014-198 & 12 &  94.96-117.23 & 134.37-136.76 \\
    \hline
    ISS\textunderscore213RI\textunderscore BMOVIE001\textunderscore PRIME & N1804922788-N1804944201 &  2015-072 & 5 &  72.71-67.14 & 82.86-83.51 \\
    \hline
    ISS\textunderscore231RI\textunderscore CDMOVIE001\textunderscore PRIME & N1832677373-N1832690125 &  2016-028 & 3 &  78.22-76.485 & 86.15-86.07 \\
    \hline
    ISS\textunderscore276RI\textunderscore HPMONITOR001\textunderscore PRIME & N1874437615-N187471965 &  2017-145 & 9 & 144.19-141.58 & 81.68-79.25 \\
    \hline
    \end{tabular}%
    }
\end{center}
\end{table*}

Figure~\ref{datpars} shows the distribution of these observations as functions of phase angle and time. Note that there is a large gap in the data between late 2009 and late 2012. This corresponds to an extended period of time when the Cassini spacecraft remained close to the planet's equator plane, and so could not easily image the main rings. Also, note that while prior to 2010 a fair fraction of the data points are isolated observations,  after 2012  virtually all the images of the ringlet came from  movie sequences where the camera stared at the Cassini Division for a significant fraction of an orbital period. These sequences are particularly useful for investigating longitudinal variations in the ringlet's brightness and so Table~\ref{obstab} summarizes the properties of several of these observations. 

All these images were calibrated using the CISSCAL routines to remove instrumental backgrounds, apply flat-fields, and convert the measured data to $I/F$, a standard unit of reflectance that is unity for a perfect Lambertian surface  viewed and illuminated at normal incidence \citep{Porco04, West10}. These images are also geometrically navigated using the appropriate SPICE kernels \citep{Acton96}. The image pointing was refined based both on the positions of stars within the field of view and the location of the outer edge of the Jeffries Gap. This geometric information was also used to determine the phase, incidence and emission angles at the ring.

Since the radial structure of the ringlet did not obviously change across the limited range of longitudes visible in a single image, we derived a radial brightness profile from each image by averaging over a range of longitudes for each image. These profiles of observed $I/F$ were then converted to  ``normal $I/F$'' $= \mu I/F$, where $\mu$ is the cosine of the emission angle. For a low optical depth ringlet like the Charming Ringlet, this quantity should be independent of emission angle and so is a more useful quantity for comparing observations. 

\begin{figure*}
\centerline{\resizebox{6in}{!}{\includegraphics{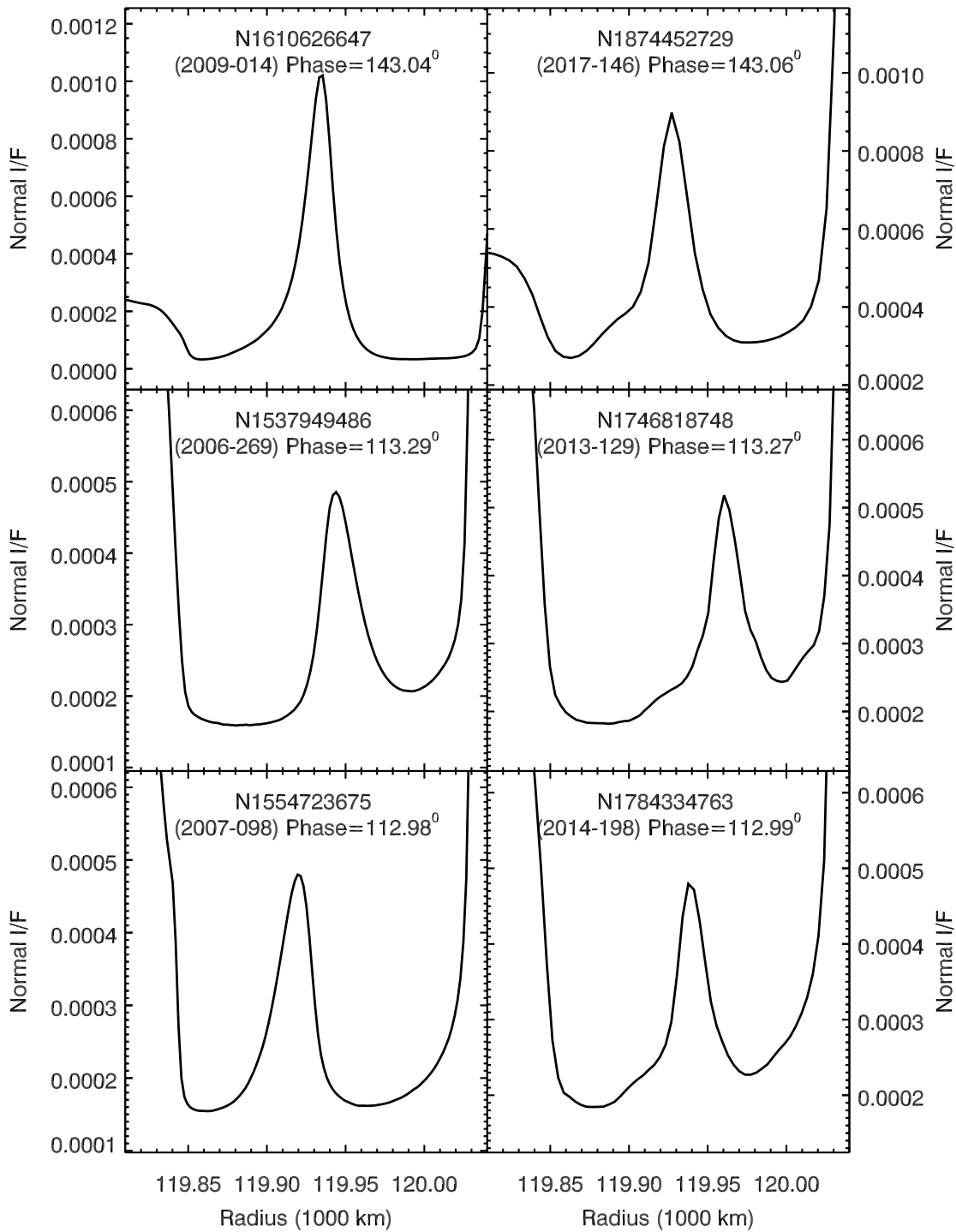}}}
\caption{Examples of ringlet profiles obtained before and after the formation of the shelf. Each of the three pairs of images were obtained at nearly the same phase angle, and so should be directly comparable to each other. {Statistical uncertainties in the brightness profiles are typically between $1\times10^{-6}$ and $2\times10^{-6}$ and so are comparable to the line width.} In the data taken in 2006-2007, the location of the ringlet varies, and shows some asymmetry in its overall shape, but still appears as a simple peak. By contrast, all the profiles obtained in 2013-2017 show a clear shelf on their inner flank. Note also that the radial extent of this shelf does not seem to change much as the position of the ringlet changes.}
\label{profsamp}
\end{figure*}

Figure~\ref{profsamp} shows some example radial profiles derived from images obtained at similar phase angles before and after the shelf appeared. The shelf can be clearly seen on the inner flank of the three images obtained after 2012, but is not present on any of the profiles taken before 2010. These data also show that the shelf is a rather subtle feature, and that the ringlet itself can have an asymmetric shape. These aspects of the ring's structure informed how we analyzed these profiles. 

\section{Survey of ringlet properties over the course of the Cassini Mission}
\label{history}

While the shelf can be seen in all the post-2010 observations in Figure~\ref{profsamp}, {quantifying its absolute brightness for all the observations is challenging because the shelf is a rather subtle feature that needs to be detangled from background trends from both the ringlet and stray light from both edges of the gap.}  For this reason, our initial investigation of these data instead focused on the total integrated brightness of the ringlet plus shelf, as well as the locations of both the ringlet and the shelf. These parameters all provide important information about what happened to the ringlet between 2010 and 2012.

\begin{figure*}
\centerline{\resizebox{6in}{!}{\includegraphics{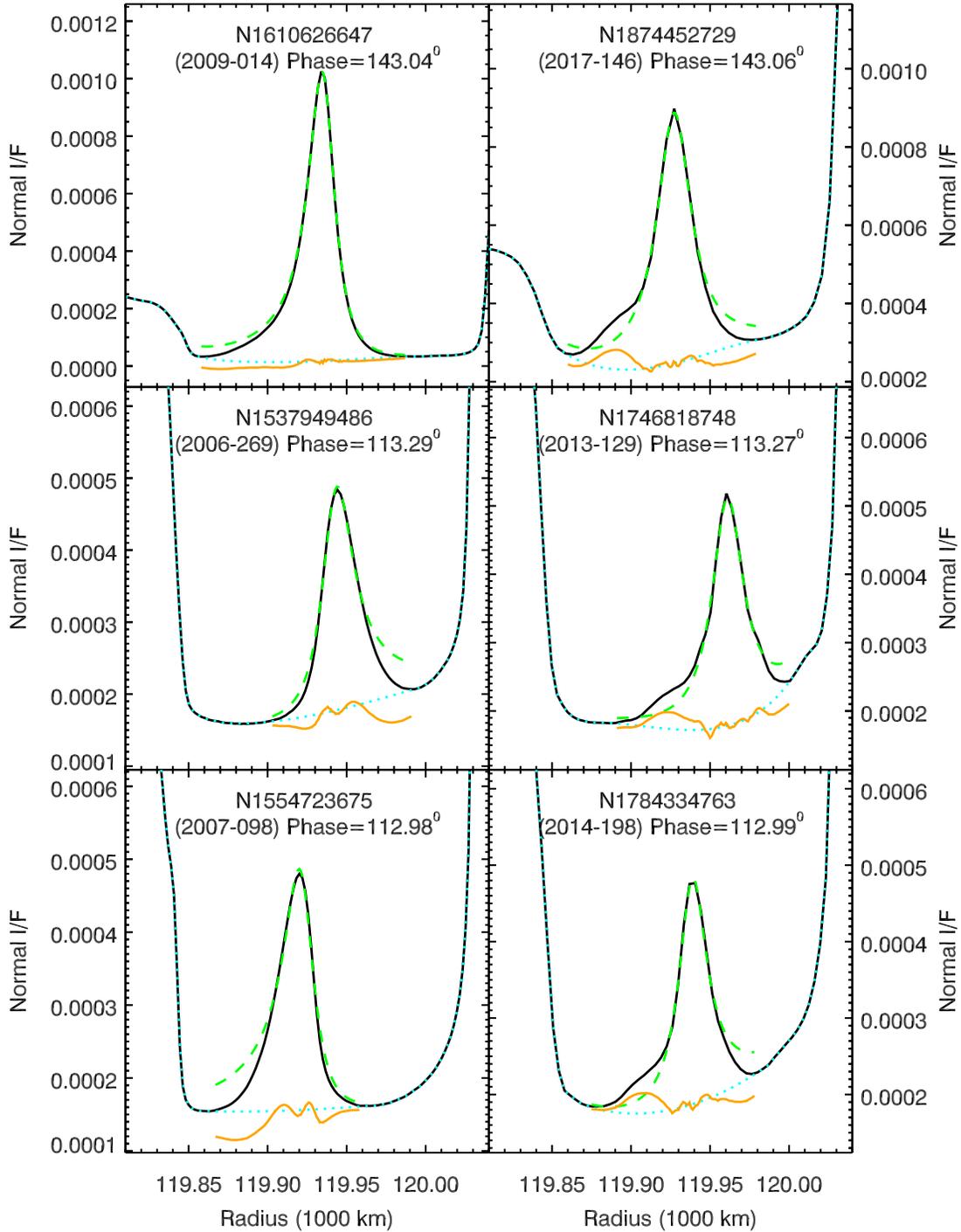}}}
\caption{Illustrations of how the signal from the shelf and ringlet are isolated. The six panels show the same radial brightness profiles of the ringlet shown in Figure~\ref{profsamp} as black solid lines. The blue dotted lines show the background signal interpolated from the region surrounding the ringlet, while the green dashed line shows the best asymmetric Lorentzian model to the ringlet's signal above this background. The orange line shows the residuals to this model relative to the background model. Note that for all the post-2010 profiles these residuals show a positive peak due to the shelf, while this feature is absent from the earlier profiles without this shelf.}
\label{profproc}
\end{figure*}

The steps we used to determine these parameters for each profile are illustrated in Figure~\ref{profproc}. First, we isolate the signal from the ringlet and the shelf from background trends associated with the surrounding ring material, so that we could determine the ringlet's total brightness. Second, we fit the background-subtracted ringlet signal to an asymmetric Lorentzian model  in order to quantify the location and shape of the ringlet, and compute the residuals to this model. The background-subtracted residual signal largely isolates the signal from the shelf, enabling the location of this feature to be quantified. Details of these procedures are provided in the following subsections. Section~\ref{brightness} focuses on the resulting estimates of the ringlet's overall brightness, which shows no dramatic changes over the course of the Cassini mission, while Section~\ref{orbit} focuses on the ringlet's orbital properties, which do shift between 2010 and 2012. 

\subsection{No evidence for large-scale changes in the ringlet's particle content}
\label{brightness}

In order to properly quantify the brightness and shape of the ringlet, we first need to isolate the ringlet signal from the backgrounds associated with nearby ring material. To accomplish this we take each profile and find the brightness peak between radii 119,900 km and 119,930 km. We then find the radii with the minimum brightness on either side of this peak and define these as $r_{min}$ and $r_{max}$. We then take the data outside the region between $r_{min}$ and $r_{max}$ and interpolate the observed brightness trends in those regions into the region between $r_{min}$ and $r_{max}$ using a cubic spline interpolation on the log-transformed data. Examples of the estimated background trends are shown as the blue dotted lines in Figure~\ref{profproc}.

In order to verify that this method properly isolates the signal from the ringlet, as well as check for possible changes in the overall brightness of the ringlet over time, we compute the radially integrated brightness of the ringlet plus shelf from each background-subtracted profile to obtain a quantity known as the normal equivalent width:

\begin{equation}
{\rm NEW} =\int \mu(I/F -I/F_{\rm back}) dr.
\end{equation}

\noindent This quantity should be independent of resolution,  emission and incidence angle, but should vary with phase angle (since the particles are more efficient at scattering light in the forward direction) and potentially time (if the change in the ringlet's structure also changed the amount of material in the ringlet). {Note that the uncertainties in these parameters are dominated by systematic errors in the background levels, and so cannot be reliably determined {\em a priori}. Instead, the uncertainties in these parameters are estimated based on the scatter in measurements obtained at similar viewing geometries.}

\begin{figure}
\resizebox{3.4in}{!}{\includegraphics{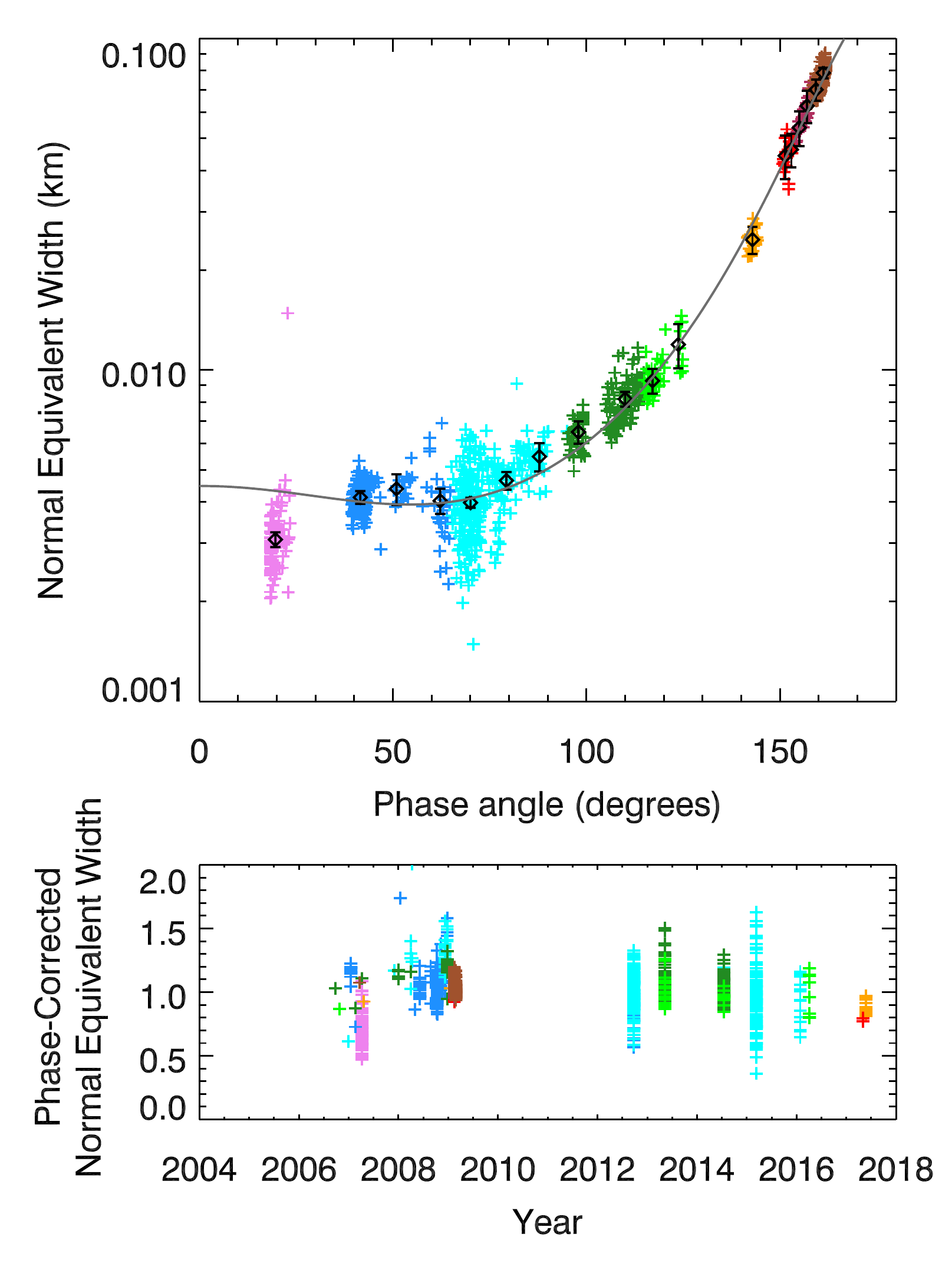}}
\caption{The radially-integrated brightness of the ringlet. The top panel shows the measured Normal Equivalent Width versus phase angle for the available images of the Charming ringlet.   {These NEW values are also provided in Table~\ref{postab}.} Colored points are individual measurements, with colors indicating phase angles. The points with error bars are average brightness values for each phase angle range (see Table~\ref{phasedat}), and the solid curve is a two-component Henyey-Greenstein function (see text). The bottom panel shows the NEW values, normalized by the best-fit phase function as a function of time (still color coded by phase angle). {This plot shows that there is no obvious shift in the ringlet's typical brightness around 2011.}}
\label{charmphase}
\end{figure}

Figure~\ref{charmphase} shows the resulting estimates of the ringlet's integrated brightness as a function of phase angle and time, {and Table~\ref{phasedat} provides the average NEW values within different phase angle bins, with uncertainties based on the observed scatter in the measurements.} While there is a dispersion of the brightness estimates around the mean trend, this dispersion is reasonably small ($<20$\%) at phase angles above  90$^\circ$. At lower phase angles the dispersion is larger (closer to 50\%), probably because the ringlet is much fainter in those viewing geometries {making the NEW estimates more sensitive to small errors in the estimated background trends.} Even so, it is important to note that the mean brightness level at phase angles below $90^\circ$ is not a strong function of phase angle, even though the brightness of the nearby ring material decreases significantly between 20$^\circ$ and 90$^\circ$ phase.  These results indicate that the above algorithms are isolating the desired ringlet signal relatively well. As mentioned above, there were 13 images that yielded integrated brightness estimates well off this trend, with NEW values above 100 meters or below 0.1 meters. These aberrant observations were excluded from the rest of this analysis.

\begin{table}
\caption{Average integrated brightness measurements of the Charming Ringlet.}
\label{phasedat}
\hspace{-.5in}\resizebox{3.5in}{!}{\begin{tabular}{|c|c|c|} \hline
Phase Range  & Mean Phase  & NEW \\ (degrees) & (degrees) & (meters) \\ \hline
      15-25 &     19.7 &    $3.12\pm 0.16$\\
      35-45 &    41.6 &      $4.14\pm    0.18$\\
      45-55 &     50.9 &      $4.39\pm     0.47$\\
      55-65 &     62.2 &      $4.10\pm     0.36$\\
      65-75 &     70.1 &      $4.01\pm     0.14$\\
      75-85 &     79.3 &      $4.70\pm     0.29$\\
      85-95 &     87.9 &      $5.56 \pm     0.55$\\
      95-100 &     98.0 &      $6.49 \pm     0.50$\\
     105-115 &     110.0 &      $8.17 \pm     0.41$\\
     115-120 &     117.1 &      $9.23 \pm    0.79$\\
     120-125 &     123.7 &      $11.4 \pm     1.7$\\
     140-145 &     142.9 &      $24.2 \pm     2.3$\\
     150-152 &     151.3 &      $43.4 \pm     6.5$\\
     152-154 &     152.3 &      $45.9 \pm     5.4$\\
     154-156 &     155.0 &      $53.2 \pm     6.5$\\
     156 -158 &     157.0 &      $61.5 \pm      6.9$\\
     158-160  &    159.3 &      $69.5 \pm    5.1$\\
     160-162  &    161.2 &      $78.3 \pm     3.0$\\ \hline
\end{tabular}}
\end{table}

To check whether there are subtle temporal variations in the integrated brightness measurements, we fit the integrated brightness data to a two-component Henyey-Greenstein function:

{
\begin{equation}
\begin{split}
NEW =  & C\frac{1}{4\pi}\left( \frac{w_1(1-g_1^2)}{(1-g_1^2-2g_1\cos\theta)^{3/2}}+\right. \\
& \left.\frac{(1-w_1)(1-g_2^2)}{(1-g_2^2-2g_2\cos\theta)^{3/2}} \right).\\
\end{split}
\end{equation}}

\noindent with fit parameters $C$, $w_1$, $g_1$ and $g_2$. The best-fit values for these quantities were {$C=0.16$ km}, $w_1=0.89$, $g_1=0.643$ and $g_2=-0.247$. This curve is shown on Figure~\ref{charmphase}, and it reproduces the observed trends between phase angles at 30$^\circ$ and $160^\circ$ quite well. The most notable issue is that the data at $20^\circ$ phase falls below the model trend, but even in this case the model is only about 20\% above the data. {This discrepancy most likely represents a limitation of the photometric model. For example, other dusty rings required a three-component Henyey-Greenstein function to reproduce the observed phase curve \citep{HS15}. Alternatively, it could indicate that our background-subtraction algorithm is starting the remove some of the real ringlet signal in cases where the surrounding ring material is sufficiently bright. In principle, we could fit these data to a more complex model and/or further refine our background-subtraction procedures to address this issue. However, in practice the only observations obtained at such low phase angles happened before the formation of the shelf, so we decided it was not worth further complicating the model/analysis to fit these data for this particular study.} 

Dividing the observed brightness data by the above phase curve yields a phase-corrected NEW that should be around one for all the observations. The bottom panel in Figure~\ref{charmphase} plots these phase-corrected brightness estimates as function of time. These data do not show strong evidence for any obvious changes in the ringlet's overall brightness over the course of the Cassini mission. {More specifically, if we compute the mean phase corrected brightness values for all phase angles above 30$^\circ$ from the observations made before and after the shelf's appearance between 2010 and 2012, we find values of $1.048\pm0.005$ and $0.980\pm0.008$ (uncertainties are based on the observed scatter in the measurements). The brightness of the ringlet therefore changed by less than 10\% between these two time periods.}  This not only confirms that these background-removal procedures are robust, but also indicates that the formation of the shelf was not associated with a major change in either the number density or typical size of the visible particles in this ringlet.

\subsection{Evidence for  changes in the ringlet's orbital properties}
\label{orbit}

In addition to changes in the ringlet's particle content, we also wanted to search for variations in the ringlet's orbital properties. We therefore fit the background-subtracted brightness profile to an appropriate functional form to estimate the ringlet's location. After some experimentation, we found that fitting the peak of the ringlet's brightness profile to an asymmetric Lorentzian function both fit the variable shape of the main ringlet reasonably well and allowed the signal from the shelf to be isolated from the fit residuals. We therefore fit the background-subtracted brightness data to the following functional form \citep{SB08}:

\begin{equation}
\mu I/F = \frac{2A/\pi \gamma}{1+4(r-r_0)^2/\gamma^2}+B
\end{equation}

\noindent where $A, B$ and $r_0$ are constants, but the width $\gamma$ is a function of radius:

\begin{equation}
\gamma =\frac{2\gamma_0}{1+e^{\alpha(r-r_0)}}
\end{equation}

\noindent where $\gamma_0$ and $\alpha$ are both constants, with $\alpha$ being the parameter that quantifies the asymmetry of the curve.  The fit is performed using the {\tt mpfitfun} IDL program \citep{Markwardt12},  and considers the background-subtracted profile wherever the brightness is over 1/3 the peak brightness, which helped to ensure that the shelf did not bias the fit. 

The green dashed lines in Figure~\ref{profproc} show examples of the best-fit functions for the selected profiles, with the background signal added back in to facilitate comparisons with the data. This function clearly fits the shape of the curve very well close to the peak, and so the fit parameter $r_0$ should provide a robust and reliable estimate of the peak location. However, it is also important to note that the asymmetric Lorentzian form does not perfectly reproduce the trends seen further from the peak. Alternative fitting functions, including  an asymmetric Gaussian, did not significantly improve these aspects of the fit. While such issues should not systematically affect the estimates of the peak locations, they do imply that estimates of the parameter uncertainties from the fit are not reliable, and so we will instead estimate the uncertainties in these parameters based on the scatter in the data.

Furthermore, the signal from the shelf is still visible as a bump in the residuals from the background-subtracted brightness data (shown as a solid orange curves in Figure~\ref{profproc}, again with the background brightness added back in to facilitate comparisons). These residuals therefore capture the signal from the shelf. For the data obtained prior to 2010, the residuals interior to the peak are either flat or monotonic, consistent with the lack of an observable shelf. By contrast, the residuals for the profiles obtained from 2012 on all show a broad peak interior to the ringlet's peak brightness whose shape and amplitude are consistent with the shelf. For this particular analysis, we are most interested in the location of the shelf, which we estimate as the location of the peak in the residual profile interior to $r_0-\gamma_0$. 

\citet{Hedman10} showed that this ringlet had both an eccentricity and an inclination, and that the eccentricity at least had two components, a ``forced" component due to solar radiation pressure and a ``free'' component that precessed around the planet at roughly the expected rate due to Saturn's non-spherical shape. Together, these aspects of the ringlet can cause the apparent position of the ringlet to shift back and forth by over 20 km. Initial investigations of all the observations indicated that the \citet{Hedman10} models for the ringlet's orbital properties were not accurately predicting the position of the ringlet in the more recent data.

In order to investigate these apparent discrepancies and to  ensure that they were not due to small pointing errors, we estimated the position of the Laplace Gap's inner edge as the point of maximum slope interior to the ringlet, and compared this to a model of its expected position that included the mean radius and eccentricity given by \citet{French16}.\footnote{The inclination and structure forced by the 2:1 Resonance with Mimas are not included in these calculations because they would only produce sub-kilometer variations in the apparent edge position.}  The differences between the observed and expected positions had a standard deviation of around 4 km, with some outliers as large as 40 km. We therefore applied a constant offset to each radial profile to bring the observed edge position into agreement with these predictions.   {These corrected values for the peak and shelf positions are provided in Table~\ref{postab}.}

Even after these corrections, the ringlet's peak location still failed to match predictions in observations made after 2010. In order to quantify these changes in the ringlet's location, we consider the following simplified model of the ringlet's radial location $r$ as a function of longitude $\lambda$ and time $t$:

\begin{equation}
r=a-ae(t)\cos(\lambda-\lambda_\Sun-\varpi'(t)),
\end{equation}

\noindent where $\lambda_\Sun$ is the sub-solar longitude, $a$ is the (assumed constant) ringlet semi-major axis, while the eccentricty $e$ and pericenter location relative to the sub-siolar longitude $\varpi'$ are both implicit functions of time given by the following relationships:

\begin{equation}
ae\cos\varpi' = ae_f\cos\varpi'_f + ae_\ell\cos(\varpi'_\ell +\dot{\varpi}'_\ell t),
\end{equation}
\begin{equation}
ae\sin\varpi' = ae_f\sin\varpi'_f + ae_\ell\sin(\varpi'_\ell +\dot{\varpi}'_\ell t),
\end{equation}

\noindent where $ae_f, ae_\ell, \varpi'_f, \varpi'_\ell$ and $\dot{\varpi}'_\ell$ are all constants, with $ae_f$ representing the eccentricity forced by solar radiation pressure and $ae_\ell$ representing the free component of the eccentricity that precesses around the planet at a rate determined by the planet's higher order gravitational field \citep{Hedman10}.  Prior investigations of this ringlet found that the ringlet also had a variable inclination {that caused the vertical position of the ring to be displaced by up to 3 km and so}  could also influence the ringlet's apparent radial position \citep{Hedman10}. However, these effects could not be robustly detected in these observations. Hence, for the sake of simplicity we do not include inclination in this analysis.

The \citet{Hedman10} model of this ringlet's shape was based on a limited number of observations that covered a range of longitudes in a short period of time, allowing the ringlet's eccentricity and pericenter location to be estimated at several different times before estimating parameters like $ae_\ell$ or $ae_f$. Fits to these data indicated that $\varpi'_f \simeq 180^\circ$ and $\dot{\varpi}'_\ell \simeq 4.68^\circ$/day, values that are consistent with theoretical expectations for a forced eccentricity induced by solar radiation pressure and the free precession rate for particles orbiting in the vicinity of the ringlet (after correcting for the apparent motion of the Sun). Also, the two components of the eccentricity were found to have the  amplitudes $ae_f= 17.0\pm0.5$ km and  $ae_\ell = 7.9\pm0.4$ km. 

Here we are considering a much more heterogeneous data set and so we instead  evaluate the $rms$ difference between the observed and predicted ringlet positions for a range of orbital parameters. In order to keep the parameter space manageable, we  assume that $\varpi'_f=180^\circ$ and $\dot{\varpi}'_\ell=4.68^\circ$/day, so that we only have to consider the parameters $ae_f$, $ae_\ell$ and $\varpi'_\ell$ (the mean radius $a$ being a constant offset that falls out of the $rms$ calculation). We evaluated the $rms$ misfit for the data taken before and after 2010 for an array of $ae_f$, $ae_\ell$ and $\varpi'_\ell$ values sampled every 1 km, 1 km and 10$^\circ$, respectively.  

\begin{figure}
\centerline{\resizebox{3.5in}{!}{\includegraphics{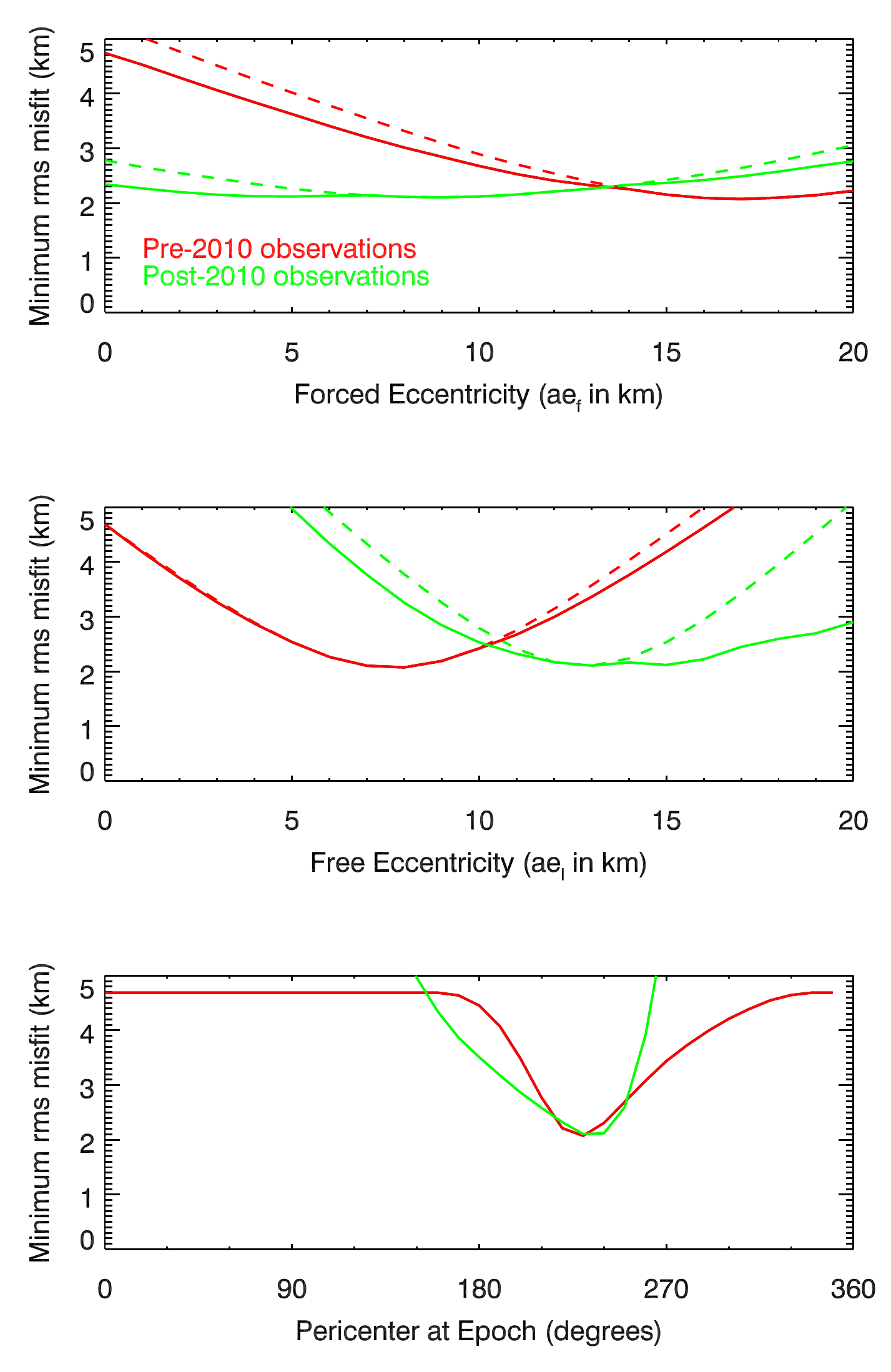}}}
\caption{Changes in the best-fit orbit parameters for the ringlet over the course of the Cassini mission. Each panel shows the $rms$ misfit between the observed ringlet peak locations and the predicted values for a three-parameter model of the ringlet's shape assuming a free precession rate $\dot{\varpi}'_\ell=4.68^\circ$/day and forced pericenter location $\varpi'_f=180^\circ$. The curves for the free pericenter location at each give the minimum $rms$ misfit over all possible values of the free and forced eccentricity, both of which show a clear minimum at 230$^\circ$. The  solid curves for the forced and free eccentricities show the minimum $rms$ misfit assuming a pericenter at epoch of 230$^\circ$, while the dashed lines show the minimum $rms$ misfit allowing the pericenter position to float. These plots show clear evidence that the ringlet's free eccentricity increased substantially sometime between 2010 and 2012. }
\label{peakfit}
\end{figure}

\begin{figure}
\centerline{\resizebox{3.5in}{!}{\includegraphics{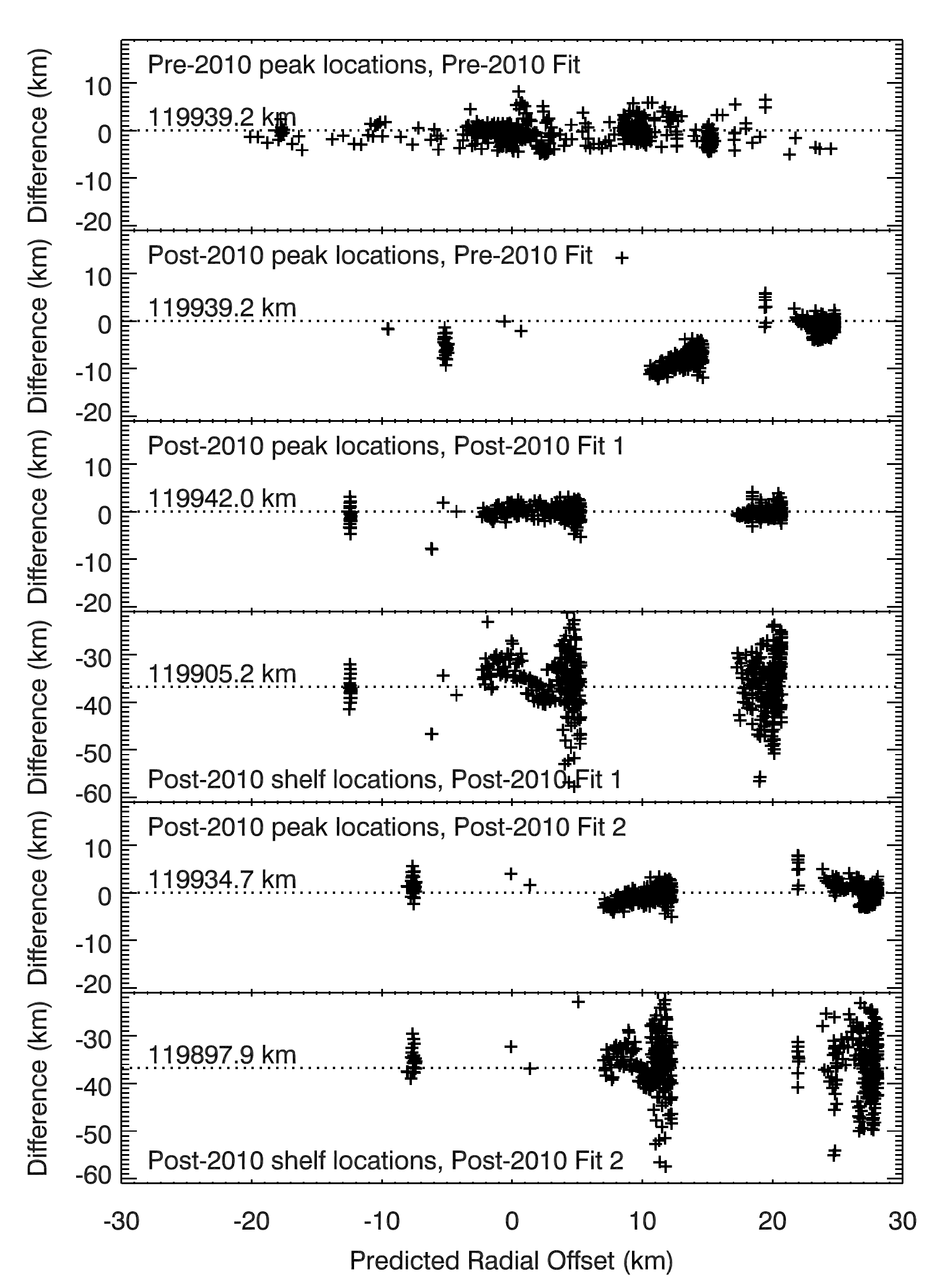}}}
\caption{Orbital changes in the ringlet. The top two panels show the differences between the observed and predicted ringlet positions assuming the best-fit pre-2010 model as functions of the predicted radial offset from the best-fit semi-major axis. The top panel shows the data obtained before 2010, which are scattered around the zero line. By contrast, the data obtained after 2010 show large systematic differences indicating that this model for the ringlet's position in no longer adequate. The lower four panels show the differences between the predicted and observed positions of the ringlet and the shelf for the two different models of the ringlet's location given in Table~\ref{orbtab}. Fit 1 is the best-fit overall model, while Fit 2 assumes $ae_f=17$ km. Both models show a much tighter dispersion around zero, and in both cases the shelf location is consistently about 37 km interior to the peak ringlet signal.}
\label{peakpos}
\end{figure}

Figure~\ref{peakfit} shows the minimum $rms$ misfit as functions of the three parameters. For the observations obtained before 2010, we find the best-fit (i.e. minimum $rms$ misfit) solution has $ae_f \simeq 17$ km, $ae_\ell \simeq 8$ km and $\varpi'_\ell \simeq 230^\circ$, which are consistent with prior results. However, for the observations made after 2010 this solution is not the one that gives the minimum $rms$ misfit. These later observations still prefer  $\varpi'_\ell\simeq 230^\circ$, but now the minimum $rms$ misfit occurs where { $ae_f\simeq 7$ km. and  $ae_\ell \simeq 14$ km} While the two minima in $ae_f$ are fairly broad, the minima in $ae_\ell$ are narrower and suggest that the free eccentricity increased sometime between 2010 and 2012. 

\begin{table*}[tb]
\caption{Orbital parameters for the ringlet}
\label{orbtab}
\begin{tabular}{l c c c c c c} \hline
Case & $ae_f$ (km) & $\varpi'_f$ (deg) &  $ae_\ell$ (km) & $\varpi'_\ell$ (deg) & $\dot{\varpi}'_\ell$ (deg/day) & $a$  (km) \\ \hline 
Hedman et al. 2010 Model 1 & 17.0$\pm$0.5 & 180$^a$ & 7.9$\pm$0.4 & 230$\pm$3 & 4.66$\pm$0.01 & --- \\ 
Pre-2010 Fit & 17.0$\pm$1.6 & 180$^a$ &  7.7$\pm$0.8 & 228$\pm$5 & 4.68$^a$ & 119939.20$\pm$0.26 \\
Post-2010 Fit 1 & 7.1$\pm$3.7 & 180$^a$ &  13.7$\pm$1.6 & 234$\pm$6 & 4.68$^a$ & 119942.31$\pm$0.10 \\
Post-2010 Fit 2 & 17.0$^a$ & 180$^a$ &  11.1$\pm$0.8 & 219$\pm$7 & 4.68$^a$ & 119933.95$\pm$0.12 \\
\hline
\end{tabular}

All fits use an Epoch time that corresponds to UTC of 2007-099T22:19:10TBD  or  229429225.185 seconds ephemeris time.

$^a$ parameter held fixed in the fit.

\end{table*}

Assuming the minimum $rms$ misfit provides a reasonable estimate of the statistical uncertainties in the position measurements, we can translate the profiles shown in Figure~\ref{peakfit} into $\chi^2$ statistics and then compute relative probabilities that can be fit to gaussians in order to estimate both the best-fit parameters and their uncertainties. These numbers are provided in Table~\ref{orbtab}, along with estimates of the ring radius derived from the mean value of the residuals after removing the variations due to the ringlet's eccentricity. The parameters for the pre-2010 data are perfectly consistent with those derived by \citet{Hedman10}, albeit with slightly larger error bars. The Post-2010 best-fit solution (designated ``Fit 1'' here) is also significantly different from the Pre-2010 solution.

In addition to providing the absolute best-fit solution for the post-2010 data as ``Fit 1", Table~\ref{orbtab} also provides parameters for a ``Fit 2'' where the forced eccentricity $ae_f$ is required to have its Pre-2010 value. This is done because the forced eccentricity should only depend on the average ringlet particle size, and there is no evidence that this changed substantially over the course of the Cassini mission (see the  previous subsection). Note that even in this case $ae_\ell$ is significantly higher after 2010. Also note that the best-fit mean radius of the ringlet for this model is about 8 km interior to its value for the best-fit solution.

Figure~\ref{peakpos} further  illustrates these changes in the ringlet's orbit properties. The top panels show the differences between the observed and expected positions of the ringlet for the Pre-2010 fit model as functions of the predicted ringlet position. The pre-2010 data show a fairly random scatter around zero difference, while the post-2010 data show systematic differences of up to 10 km. The differences for the two Post-2010 fits are both much tighter, although there are clear trends in subsets of the data for Fit 2, as is to be expected given this is not the fit with the minimal $rms$ misfit. In addition, we can note that the two models predict very different values for the ringlet's radial offsets, with Fit 2 generally predicting the data fall closer to apocenter, which is consistent with Fit 2 having the lower value of $a$.

Figure~\ref{peakpos} also shows the locations of the shelf for these two fits. These positions have a considerably larger scatter because the shelf is a more subtle feature, but there is not a strong trend in the shelf position relative to the ringlet, consistent with the appearance of the profiles in Figure~\ref{profsamp}. This implies that the shelf consists of particles with roughly the same eccentricity as the ringlet, but a smaller semi-major axis. At the same time, there are hints of a trend in the shelf position within the data around 0 km in Fit 1 and 10 km in Fit 2. This suggests that there might be longitudinal variations in the shelf's structure. Both of these findings are supported by a more detailed investigation of selected movie sequences made after 2012.

\section{Longitudinal variations in the structure of the ringlet and shelf between 2013 and 2015}
\label{long}

As mentioned above, most of the data on this ringlet obtained after 2010  were movie sequences where the camera took multiple images of the ringlet over the course of a timespan comparable to the orbital period of the ring material. Detailed examinations of the different images within several of these sequences revealed subtle variations in the brightness profiles that appear to reflect longitudinal variations in the structure of both the ringlet and the shelf. 

Documenting these brightness variations is challenging because they are subtle, amounting to only 5\% of the ringlet's peak brightness, and so can easily be obscured by slight radial shifts in the ringlet's position and variations in the ringlet's overall brightness due to small changes in the observed phase angle. Hence, in order to document the longitudinal variations in the structure of this ringlet, we process the relevant brightness profiles to create aligned, background-subtracted and normalized profiles that can easily be co-added and compared. We first align the brightness profiles by interpolating  the brightness data for each profile onto a common regular grid of radii relative to the peak location estimated from the asymmetric Lorentzian fit discussed in the previous section. To improve signal-to-noise, we average together all profiles obtained within pre-determined ranges of co-rotating longitudes, which are computed for each profile assuming the material had a mean motion consistent with a particle orbiting at a specified semi-major axis. A linear background is then removed from these combined profiles based on the signal levels at the local minima on either side of the ringlet. Finally, the brightness values are normalized so that the peak brightness of the ringlet was unity. 

These procedures were applied to data from five of the movies in Table~\ref{obstab}. Specifically, we consider Rev\footnote{``Rev" designates each orbit of Cassini around Saturn} 103 SHRTMOVIE, Rev 189 BMOVIE, Rev 206 BMOVIE, Rev 213 BMOVIE and Rev 276 HPMONITOR. The first of these is the longest movie obtained prior to 2010 at high phase angles, and so provides a useful baseline for the later observations. Rev 189 BMOVIE, Rev 206 BMOVIE and Rev 276 HPMONITOR each cover roughly one orbital period of the ringlet, and so provide the clearest picture of the longitudinal variations in the ringlet after 2011. The Rev 213 BMOVIE observation is also included because it covers about half an orbit period and does preserve information about the longitudinal structure of the ringlet/shelf. The other movies in Table~\ref{obstab} were found to be too short to provide clear information about the longitudinal structure of the ringlet.

\begin{figure}
\centerline{\resizebox{3.15in}{!}{\includegraphics{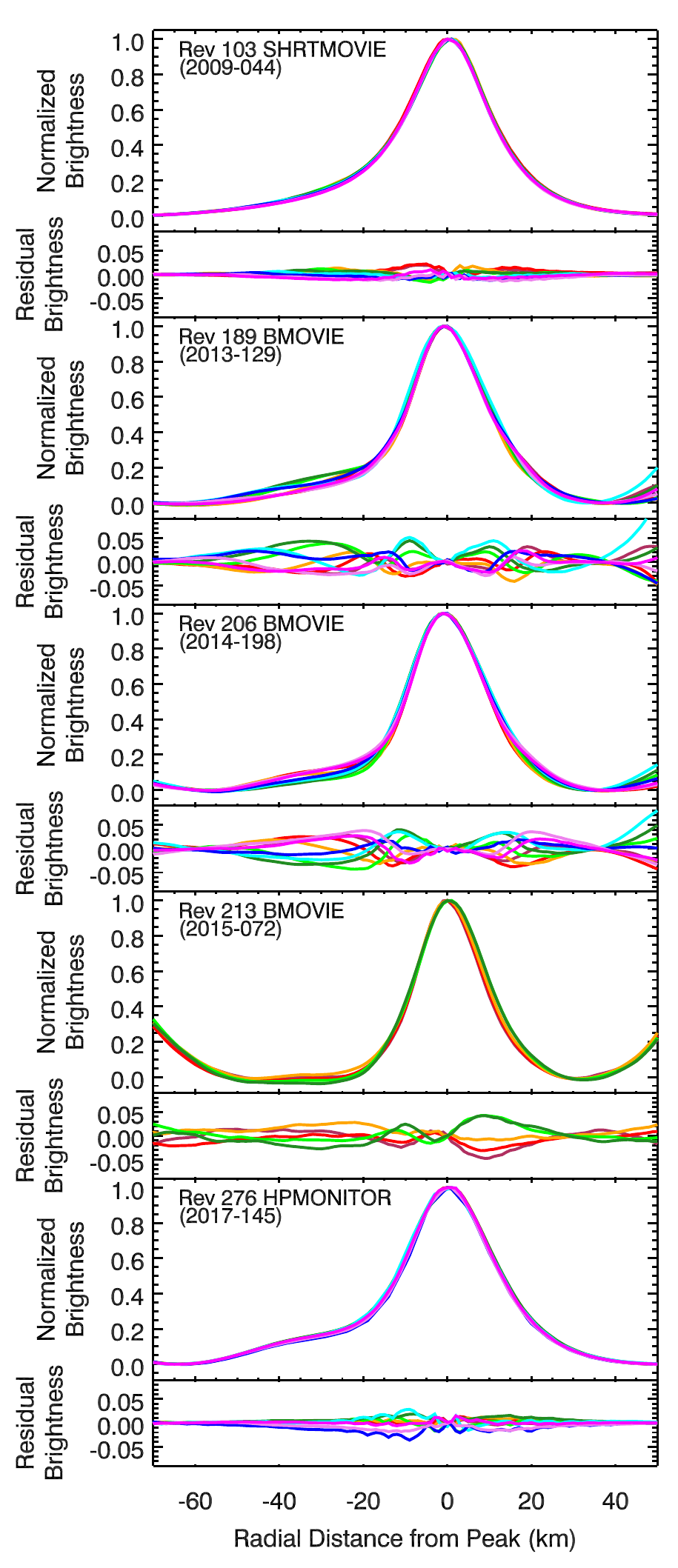}}}
\caption{Normalized, background-subtracted profiles derived from five movie sequences. For each movie, the profiles shown in  different colors correspond to different co-rotating longitudes assuming the mean motion of  736.388$^\circ$/day (corresponding to semi-major axis of 119,930 km) and an epoch time of 2011-341T00:00:00 UTC. For each movie, the top panel shows the actual profiles, while the lower panel shows the differences between each profile and the average of all profiles. {The data plotted here is provided in Table~\ref{proftab}.}}
\label{lonprofiles}
\end{figure}

\begin{figure}
\centerline{\resizebox{3.2in}{!}{\includegraphics{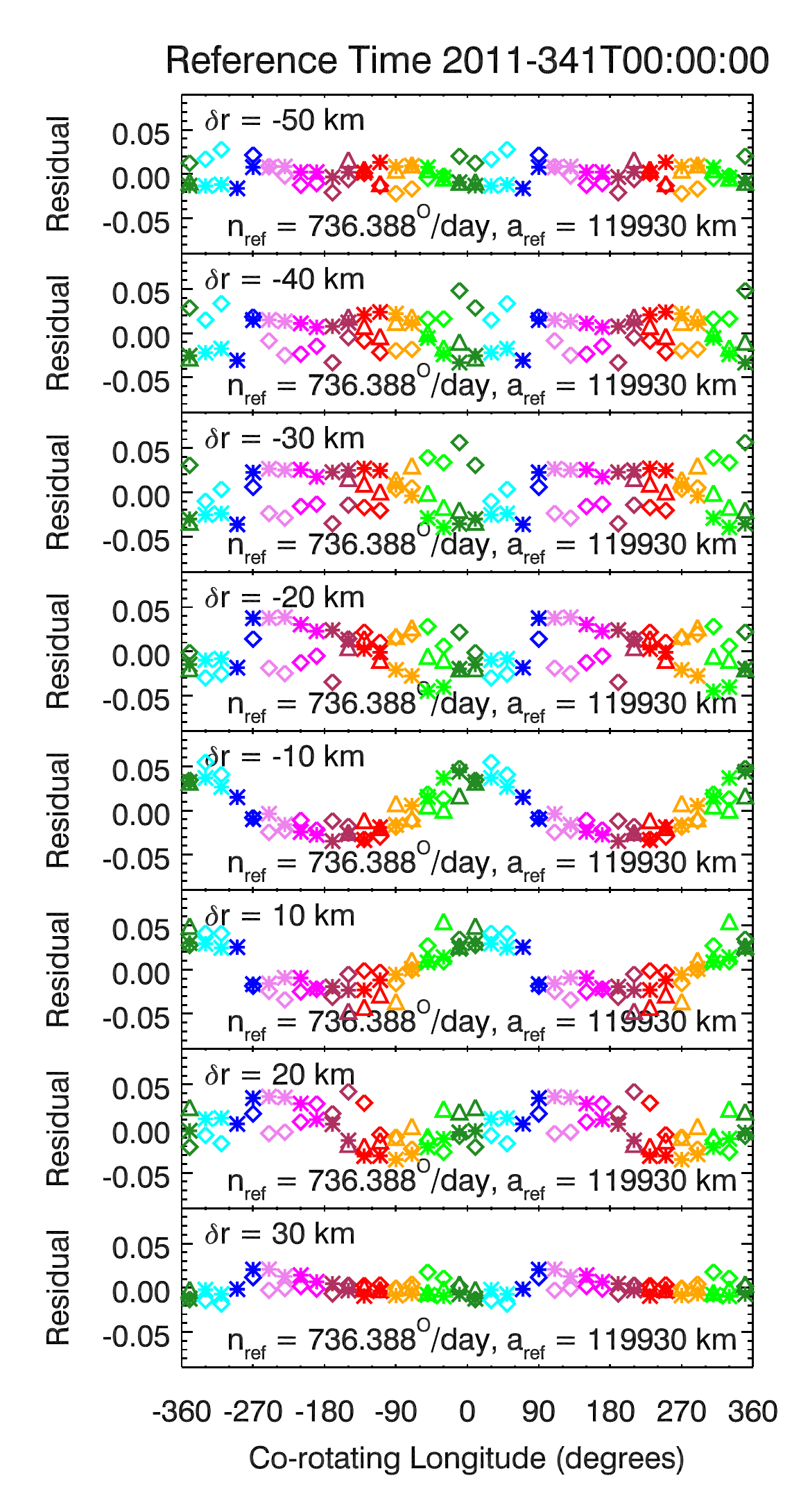}}}
\caption{Comparisons of the residual brightness variations in the ringlet as functions of co-rotating 
longitude  for the Rev 189, 206 and 213 BMOVIE sequences, shown as diamonds, stars and triangles, respectively. Each panel shows the residual brightness variations at a particular radius versus co-rotating longitude assuming a rotation rate of 736.388$^\circ$/day and an epoch time of 2011-341T00:00:00, so that the color code is the same as for Figure~\ref{lonprofiles}.  Note that the data in each panel are repeated twice for clarity. The brightness variations in the panels within $\pm10$ km of the peak are well aligned, but those further away from the peak are not well aligned for this particular rotation rate.  {The data plotted here is provided in Table~\ref{proftab}.}}
\label{lonprofiles2}
\end{figure}

\begin{figure}
\centerline{\resizebox{3.2in}{!}{\includegraphics{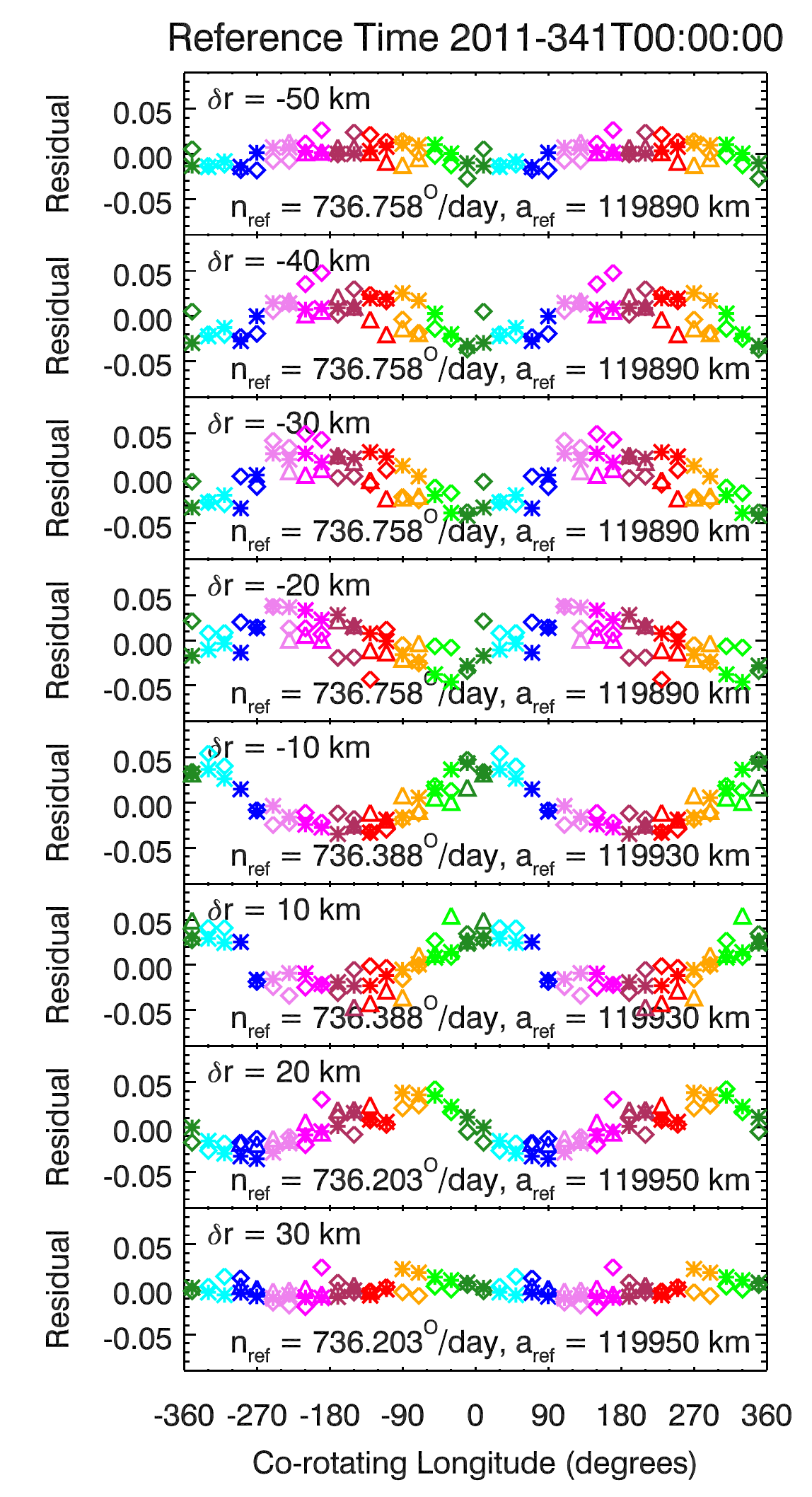}}}
\caption{Comparisons of the residual brightness variations in the ringlet as functions of co-rotating 
longitude  for the Rev 189, 206 and 213 BMOVIE sequences, shown as diamonds, stars and triangles, respectively. Each panel shows the residual brightness variations at a particular radius versus co-rotating longitude. Note that the data are repeated twice for clarity. For each movie, the profiles shown in  different colors correspond to different co-rotating longitudes assuming the mean motion provided  and an epoch time of 2011-341T00:00:00 UTC.  Note that the brightness variations in the shelf are well aligned assuming a rotation rate of around 736.388$^\circ$/day, while the variations around the ringlet peak are better fit by a rate around 736.758$^\circ$/day, and the variations exterior to the peak are aligned with a rate of 736.203$^\circ$/day.}
\label{lonprofiles3}
\end{figure}

Figure~\ref{lonprofiles} shows the resulting normalized, background-subtracted profiles derived from these five observations, along with the differences between each profile and the average of all the profiles to better show the variations. Staring with the Rev 103 SHRTMOVIE, we see that all the normalized profiles have very similar shapes, with residual differences less than 0.02 the peak brightness. This demonstrates that this ringlet was longitudinally homogeneous prior to 2010.  By contrast, the Rev 189, 206 and 213 BMOVIE sequences all show longitudinal brightness variations of order 0.05 the peak brightness across the ringlet and the shelf. Finally, the Rev 276  HPMONITOR movie appears to be relatively homogeneous, with residual brightness variations less than 0.02. 

The Rev 103 SHRTMOVIE and Rev 276 HPMONITOR observations were obtained at higher phase angles than the three BMOVIEs (over 140$^\circ$ versus  less than 120$^\circ$), so one could argue that the ringlet is just more homogeneous in higher-phase observation geometry. However, we regard this explanation as unlikely, and will instead argue that whatever event disturbed the ringlet between 2010 and 2012 produced a series of localized disturbances in the ring that rotated around the planet at different rates and therefore gradually smeared out around the ring, causing the longitudinal variations to dissipate between 2013 and 2017.

A closer look at the BMOVIE sequences reveals three distinct radial zones in the brightness variations. The innermost zone falls between -50 km and -20 km from the peak location, and corresponds to the shelf. Indeed, it appears that the profiles with the minimum brightness in this region basically do not have a shelf, while the highest brightness profiles have a shelf comparable to that seen at all longitudes in 2017. Next, there are the brightness variations within about 15 km of the peak location. These brightness variations go to zero at the peak because of how the profiles are normalized, and indicate that the width of the ringlet varies longitudinally. Finally,  around +20 km outside the peak there are variations in the ringlet brightness that are out of phase with the variations closer to the peak, which cause an inflection in the normalized brightness profiles at certain longitudes. This last feature is clear in both the Rev 189 and Rev 206 BMOVIE data, but is not clearly visible as a distinct feature in the Rev 213 BMOVIE data, most likely because in this particular observation the variations within and outside the peak happen to be aligned with each other.

It is important to note that these three brightness variations do not appear to be rotating around the planet at the same rate.  All the data shown in Figure~\ref{lonprofiles} uses the same color-code, and so the same colored curves in the different panels are at the same co-rotating longitudes assuming a mean motion of 736.388$^\circ$/day  (this rate is the expected mean motion for material at a semi-major axis of 119930 km). In this frame, the brightness variations around the peak are well aligned between the various observations, with the green/cyan-colored profiles being high and the red/pink-colored profiles being low. However, if we look at the variations in the shelf and outside the main peak, we see they are not aligned in the same way, indicating that the brightness maxima in these regions are not traveling around the planet at exactly this rate. 

These variations in the pattern alignment can also be seen in Figure~\ref{lonprofiles2}, which shows the same data in a different form. Here we show the brightness variations at particular radii as functions of co-rotating longitude (using the same color code as in Figure~\ref{lonprofiles}), but data from the different observations are shown with different symbols. In general, the longitudinal brightness variations are quasi-sinusoidal. Also, for  radii within $\pm$15 km of the peak the variations from the different observations are pretty well aligned. However, this is not the case for radii further from the peak. At $+20$ km, there are clear offsets between the three observations, and for radii between $-20$ and $-50$ km in the shelf, the 189 and 206 BMOVIE data show peak brightnesses at points roughly 180$^\circ$ apart. 

After some experimentation, we found that the brightness variations in the shelf are better aligned if we assume a rate of 736.758$^\circ$/day, while the  variations outside the peak are well aligned assuming a rate of 736.203$^\circ$/day. These rates correspond to semi-major axes of 119,890 km and 119,950 km, or about -40 km and +20 km from the semi-major axis that aligns the variations within the peak. These numbers are therefore consistent with the observed radial positions of these features. Figure~\ref{lonprofiles3} shows the brightness variations aligned using these different rotation rates. The variations at $+20$ km are clearly much better aligned using the slower rotation rate, and the variations around $-30$ and $-40$ km are also reasonably well aligned with the faster rate. Note the situation at -20 km is less clear, most likely because the longitudinal variations in the ringlet and the shelf are interfering with each other. Still, the data indicate the ringlet's brightness variations involve material with a range of semi-major axes and mean motions, which is consistent with the idea that these variations sheared out between 2015 and 2017. 


\section{Discussion}
\label{discussion}

The above analyses provide clear evidence that something happened to this dusty ringlet sometime between 2010 and 2012 that both produced a shelf of material interior the ringlet, and altered the average orbital properties of the particles in the ringlet. Determining what could have caused these changes is challenging because we do not know exactly when these changes were initiated. Since  the Rev 172 MOONLETCD observation in 2012 already shows the ringlet's position systematically deviates from the pre-2010 model, we know that whatever disturbed the ring started before this. Unfortunately, Cassini was orbiting close to Saturn's ring-plane for most of this time period, and so there are virtually no images that could directly document what happened to the ring between 2010 and 2012. Instead, we have to use the available data to indirectly constrain what might have happened to the ringlet.

When evaluating possible mechanisms for changing the shape and orbit of this ringlet, it is important to remember that the particles in the ringlet and the shelf are very small. The magnitude of the heliotropic eccentricity before 2010 implies that the average effective particle radius in the ringlet is around 20 $\mu$m \citep{Hedman10}, which is consistent with the observed strength of the Chirstainsen feature in occultations of this ringlet \citep{Hedman11f}. The lack of an obvious change in the ringlet's phase function over the course of the Cassini mission suggests that the changes in the ringlet's orbital properties did not alter the ringlet's particle size distribution much. Furthermore, the shelf is about 10\% the ringlet's peak brightness at multiple phase angles (see Figure~\ref{profsamp}), so there does not also appear to be a major difference in the particle size distribution between the shelf and the ringlet after 2012. Hence we may assume here that the typical particle in the ringlet and the shelf has a radius of roughly 20 $\mu$m. 

In principle, there are a wide variety of processes that could perturb the orbits of the small particles found in this ringlet, including solar radiation pressure, Poynting-Robertson drag, changes in the local electromagnetic fields and/or plasma environment, and collisions with interplanetary debris. However, in practice, collisions with interplanetary debris is the option that seems most consistent with the longitudinal brightness variations seen between 2012 and 2016 and the orbital properties of the particles in the shelf. 

The longitudinal brightness variations indicate that this process did not affect all parts of the ringlet equally. This argues against the idea that these changes were produced by a global, uniform process, and instead  suggests that the perturbation was a discrete event that was sufficiently  localized in time and space to affect certain longitudes more than others. This disfavors phenomena like solar radiation pressure and Poynting-Robertson drag because they depend on solar flux, which is unlikely to vary on sufficiently short timescales to only affect parts of the ringlet. Also, while changes in the global configuration of the electric and magnetic fields have been observed  \citep{Andri14, Provan14}, such large-scale variations might have trouble producing sufficiently  local perturbations within this ringlet. 

 Meanwhile, even tough both the mean eccentricity and semi-major axis of the ringlet changed during this event, the largest orbital changes were experienced by the material that generated the shelf. Both the radial locations of the shelf and the rotation rates of its longitudinal asymmetries indicate that the particles in the shelf are on orbits with semi-major axes around 40 km closer to Saturn than the ringlet. If we make the reasonable assumption that the shelf material originally came from the ringlet, then this finding has clear implications regarding the direction of the forces that acted on these particles. 
 
 In order for the particles' semi-major axes to shift inwards, the particles need to lose orbital energy, so they need to experience a force that opposed their orbital motion. This provides further evidence against the perturbation being due to sudden localized shifts in the planet's magnetic field since magnetic fields do no work. This also suggests that the disturbance was not due to a sudden change in the plasma environment. While one could imagine some event suddenly releasing a large amount of plasma into a small part of the ringlet, that plasma would naturally be picked up by the magnetic field and so move around the planet at Saturn's rotation rate, which is faster than the ring-particles' Keplerian orbital rate. Momentum exchange between such a plasma and the ring particles would therefore cause the ringlets to move outwards rather than inwards. 
 
Given the above challenges associated with attributing the changes in the Charming Ringlet's structure and orbit to interactions with the planet's electromagnetic field, the local plasma environment and solar radiation pressure, we are left with the possibility that the ringlet was disturbed by collisions with interplanetary debris. So long as the debris only passed through a small part of the ringlet and approached the ring from an appropriate direction, such collisions could easily produce the localized disturbances and the orbital energy reduction required by the observations. 

In principle, these collisions could have involved a small number larger objects that crashed into source bodies within the ringlet and released new material, or a larger number of tiny particles that perturbed the orbits of the dust-sized particles within the ringlet. However, given the overall brightness of the ringlet did not change much between 2010 and 2012, the latter option appears to be more likely. Furthermore, there are precedents for attributing structural changes in dusty rings to collisions with debris clouds. Collisions with debris from Shoemaker-Levy 9 appears to have created vertical spiral patterns in Jupiter's dusty rings \citep{Showalter11}, and similar collisions with cometary debris is a plausible explanation for spiral patterns in Saturn's D ring \citep{Hedman11c, Hedman15, Hedman16}. Furthermore, one of these D-ring patterns was likely formed in late 2011, so there is even evidence that a debris cloud could have been passing through the rings at roughly the right time to initiate the observed changes in the Charming Ringlet.  

However, one potential concern with this scenario emerges when we estimate the magnitude of the impulses that needed to be imparted to the particles in the Charming Ringlet in order to produce the shelf.  The standard orbital perturbation equations for near-circular orbits say that the semi-major axis of a particle evolves at a rate \citep{Burns76, Hedman18}:

\begin{equation}
\frac{\partial a}{\partial t} = 2na\frac{F_p}{F_G}
\end{equation}
\noindent where $F_p$ is the azimuthal component of the perturbing force, { $n=\sqrt{GM/a^3}\simeq736.39^\circ$/day} is the particle's mean motion, and $F_G=GMm_p/a^2=n^2am_p$ is the planet's central gravitational force on the particle, where $G$ is the gravitational constant, $M$ is the planet's mass and $m_p$ is the particle's mass. Thus the particle will undergo a semi-major axis change $\delta a$ when it receives an azimuthal impulse given by the following formula:

\begin{equation}
F_p \delta t =\frac{F_G}{2na} \delta a = \frac{1}{2} nm_p\delta a
\end{equation} 

\noindent Assuming the particles in the shelf are composed primarily of water ice with negligible porosity and have an average radius of 20 $\mu$m, we may conclude that the typical mass of these particles is around $3\times10^{-11}$ kg, and so the required impulse to produce a semi-major axis shift of 40 km is $10^{-10}$ kg m/s. 

This impulse can be delivered by a collision with a small piece interplanetary debris. So long as the impactor is much smaller than the ring particle, the impulse delivered to the ring particle will be comparable to the momentum of the impactor $m_iv_i$, where $m_i$ is the mass of the impactor and $v_i$ is the impact speed. Furthermore, for any collision between a ring particle and interplanetary debris, the impact speed will be comparable to the ring-particle's orbit speed, so we can estimate that $v_i \sim 20$ km/s. In this case, the required momentum could be delivered by a debris particle with a mass of $m_i\simeq 6\times10^{-15}$ kg, which corresponds to a solid ice grain with a radius of around 1 micron. Note that this inferred impactor mass is consistent with the prior assumption that $m_i<<m_p$.

The problem with this particular scenario is that the collision may not just change the particle's orbit, but  could also cause the particle to break apart. The standard metric for whether a collision disrupts a body is the ratio of the collision energy to the mass of the target particle $Q$, which is this case can be well approximated as:

\begin{equation}
Q \simeq \frac{1}{2}\frac{m_i}{m_p} v_i^2  \simeq \frac{1}{4}\frac{\delta a}{a} v_i^2
\end{equation}

\noindent Again assuming impact speeds of around 20 km/s, we find that the collisions required to shift the particle's semi-major axes by 40 km have $Q  \simeq 3 \times10^4$ J/kg or $3\times10^8$ erg/g.  This is close to the expected threshold energy for catastrophic disruption for solid ice grains. 

Extrapolating the disruption energy thresholds computed by \citet{BA99} down to 20 $\mu$m yields a critical disruption threshold $Q^*_D \sim 2\times10^8$ erg/g  for icy objects colliding at speeds of 3 km/s. Scaling this threshold by $v_i^{0.5}$, as recommended by \citet{Krivov18} based on work by \citet{SL09} increases this threshold slightly to around $4\times10^{8}$ erg/g. This number is also roughly consistent with the expected fragmentation threshold for ice-rich dust derived by \citet{BD95}, which is estimated to be roughly the ratio of particle's dynamic tensile strength to its mass density. Assuming mass densities of around 1 g/cm$^2$ and dynamic tensile strengths of around 20 MPa \citep{LA83}, one also obtains disruption thresholds around $2\times10^8$ erg/g. However,  it is important to note that laboratory experiments colliding centimeter-scale ice blocks generally yield orders of magnitude lower disruption thresholds than these numbers \citep[see][]{SL09}. While this may in part be because these experiments are at much lower speeds and often involve polycrystalline ice, we certainly do not have sufficient information to determine whether the collisions needed to move material into the shelf would actually destroy the particles or not. Future constraints on the strength of tiny ice grains therefore could either support or refute this particular scenario.

Despite the challenges associated with any of the above mechanisms for producing the observed shifts in the structure of the ringlet, further examination of the total required momentum input into the ring yields evidence for a connection between the changes in the ringlet and the nearly contemporaneous appearance of spiral patterns in D ring mentioned above. 

Consider a small part of the shelf that contains $\mathcal{N}$ particles per unit area, the total impulse per unit area needed to move all those particles into the shelf is

\begin{equation}
J = \mathcal{N}F_p\delta t = \frac{1}{2}\mathcal{N}m_p n\delta a
\end{equation}

\noindent For a tenuous ring consisting of particles of size $r_p$, the optical depth is $\tau = \mathcal{N} \pi r_p^2$, while the mass of such particles is $m_p=(4\pi/3)\rho_p r_p^3$, where $\rho_p$ is the the particle's mass density. We can therefore re-write this impulse density as:

\begin{equation}
J = \mathcal{N}F_p\delta t = \frac{2}{3}\tau \rho_p r_p n\delta a
\end{equation}

\noindent This expression can be generalized to a ringlet with a range of particle sizes by simply re-interpreting $r_p$ as the effective average size of the relevant particles. For this particular ringlet, we may assume  $r_p \simeq 20 \mu$m, and since all the particles are composed primarily of water ice, we may also  say  $\rho_p \simeq 1$ g/cm$^3$. Furthermore,  since the shelf is about 10\% as bright as the ringlet, which has a peak optical depth of around 0.001, we may also assume $\tau \simeq 10^{-4}$. This yields a impulse density of  around $8\times10^{-6}$ kg/m/s. 

This number is interesting because it is comparable to the impulse density required to generate the spiral pattern in the D ring in late 2011, which \citet{Hedman16} estimated was around $40\times10^{-6}$ kg/m/s. Furthermore, if interplanetary material was responsible for both structures, the impulse density into the D ring would naturally be higher due to its location deeper in Saturn's gravitational potential well. These numbers therefore provide additional support for the idea that interplanetary material might have perturbed both these  dusty rings. 

Finally, we  can further explore this potential connection between the two rings by considering what the longitudinal brightness variations in the ringlet would have looked like when the D ring was being disturbed. The date of the D-ring disturbance can be accurately estimated based on the predictable evolution of the spiral pattern that it generated, and turns out to be within a week of December 7,  2011 \citep{Hedman16}. Figure~\ref{lonprofiles3} uses this as the epoch time to define the co-rotating longitudes, so this plot shows the locations of the brightness variations in the various regions at that particular time. While the absolute longitudes of these features change rapidly as the particles orbit the planet, the relative motion of the brightness variations is much slower, and so this graph should provide a good sense of what the radial distribution of the brightness of the features would be at this particular time. Interestingly, the brightness maxima in the shelf interior to the ringlet are roughly aligned with brightness minima in the ringlet itself. This could be consistent with roughly 10\% of the material in the ringlet being thrown inwards by about 40 km. However, we caution that this could also be a chance alignment. For example, near the start of 2011 the brightness asymmetries at all radii are nearly aligned with each other. 

At this point, we cannot unambiguously prove that the disturbances to the Charming Ringlet were caused by the same event that produced a spiral pattern in the D ring. However, if the same event was responsible for disturbing both these rings, then that event may have effected the structures and orbits of other dusty rings in the Saturn system. Detailed studies of those dusty systems should therefore be a productive avenue for future work.

\vspace{1in}

\section*{Acknowledgements}

We thank the Cassini Project and Imaging Team for acquiring the data used in this analysis, as well as the Planetary Data System Ring-Moon System Node for making these data easily available. This work was supported in part by a NASA Cassini Data Analysis Program Grant  NNX15AQ67G.

\begin{table*}
\caption{Ringlet Parameters derived from the Cassini Images. See supplemental information for  a machine-readable version of this entire table.}
\label{postab}
\resizebox{6.5in}{!}{\begin{tabular}{ccccccccc}\hline
Image Name & Image Midtime & Observed & Sub-Solar  & Emission & Phase & NEW & Peak Radius & Shelf Radius \\
 & (seconds)$^a$ & Longitude (deg)$^b$ & Longitude (deg)$^b$ & Angle (deg) & Angle (deg) & (km) & (km)$^c$ & (km)$^c$ \\ \hline
        N1477740094  &       152319435.4     &           79.9       &        164.6       &         99.0        &       145.9     &        0.02809     &       119946.6   &             -1.0 \\
         N1537949486   &      212528441.0      &         194.6      &        -169.4       &         57.8       &        113.3             & 0.00874         &   119951.9        &        -1.0 \\
         N1540572432    &     215151370.6         &      146.4       &       -168.3        &       105.0      &         133.4             & 0.01345      &      119957.5         &       -1.0 \\
         N1540573152     &    215152090.6        &       146.4     &         -168.3       &        105.2       &        133.0             & 0.01351       &     119958.7        &        -1.0 \\
         N1540598112   &      215177050.5       &        132.1       &       -168.3         &      113.4      &         124.9              & 0.01093       &     119955.2         &       -1.0 \\
         N1546193738     &    220772641.1        &       221.4         &     -166.1        &       144.0           &     76.4             & 0.00273      &      119953.1    &            -1.0 \\
         N1547768529    &     222347422.4       &        135.3         &     -165.5        &        87.1        &        41.8             & 0.00466        &    119952.7       &         -1.0 \\ \hline
\end{tabular}}

$^a$ Ephemeris time in seconds past J2000 epoch

$^b$ Longitudes measured in an intertial frame in Saturn's equator plane relative to the plane's ascending node on the J2000 coordinate system.

$^c$ Positions after correcting profiles to match the predicted location of the Laplace Gap's inner edge. Shelf radius is set to -1 km prior to 2011. 
\end{table*}

\begin{table*}
\caption{Normalized, aligned ringlet profiles from selected movie sequences. See supplemental information for a  machine-readable version of this entire table}
\label{proftab}
\resizebox{6.5in}{!}{\begin{tabular}{ccccccc}\hline
Distance from & Co-rotating & \multicolumn{5}{c}{Normalized brightness for Observation}  \\
Peak (km) & Longitude$^a$(deg) & Rev 103 SHRTMOVIE & Rev 189 BMOVIE & Rev 206 BMOVIE & Rev 213 BMOVIE & Rev 276 HPMONITOR \\ \hline
               -75.0  &            -170.0      &        0.0006     &         0.0120       &       0.0713     &           -NaN         &     0.0473 \\
               -74.0    &          -170.0     &         0.0009       &       0.0098   &           0.0633       &         -NaN    &          0.0389 \\
               -73.0      &        -170.0     &         0.0013       &       0.0075    &          0.0554        &        -NaN     &         0.0306 \\
               -72.0      &        -170.0      &        0.0017      &        0.0054     &         0.0476       &         -NaN      &        0.0222 \\
               -71.0       &       -170.0      &        0.0023      &        0.0035      &        0.0409       &         -NaN       &       0.0139 \\
               -70.0      &        -170.0      &        0.0030     &         0.0017       &       0.0349       &         -NaN        &      0.0093 \\
               -69.0     &         -170.0      &        0.0038     &         0.0000       &       0.0290       &         -NaN       &       0.0070 \\
               -68.0     &         -170.0      &        0.0048     &        -0.0015        &      0.0242      &          -NaN        &      0.0047 \\
               -67.0     &         -170.0        &      0.0059    &         -0.0031        &      0.0200     &           -NaN        &      0.0023 \\
               -66.0     &         -170.0       &       0.0070   &          -0.0044        &      0.0157    &            -NaN        &      0.0000 \\
\hline
 \end{tabular}}
 
 $^a$ Assuming $a=119,930 km$ and an epoch time of 2011-341T00:00:00
 
 \end{table*}


\end{document}